# Phase-Modulated Rapid-Scanning Fluorescence-Detected Two-Dimensional Electronic Spectroscopy

*Damianos Agathangelou[1], Ariba Javed[1,2], Francesco Sessa[1], Xavier Solinas[3], Manuel Joffre[3] and Jennifer P. Ogilvie[1, 3]**

[1]Department of Physics and Biophysics, University of Michigan, 450 Church St, Ann Arbor, MI 48109 USA
[2]Department of Materials Science and Engineering, University of Michigan, 2300 Hayward St, Ann Arbor, MI 48109-2136 USA
[3]Laboratoire d'Optique et Biosciences, Ecole Polytechnique, CNRS, INSERM, Institut Polytechnique de Paris, 91128 Palaiseau, France

**\*Corresponding Author** jogilvie@umich.edu

We present a rapid-scanning approach to fluorescence-detected two-dimensional electronic spectroscopy that combines acousto-optic phase-modulation with digital lock-in detection. The approach shifts the signal detection window to suppress 1/f laser noise and enables interferometric tracking of the time delays to allow for correction of spectral phase distortions and accurate phasing of the data. This use of digital lock-in detection enables acquisition of linear and nonlinear signals of interest in a single measurement. We demonstrate the method on a laser dye, measuring the linear fluorescence excitation spectrum, as well as rephasing, non-rephasing and absorptive fluorescence-detected two-dimensional electronic spectra.

**TOC GRAPHICS**

**Introduction:**

Two-dimensional Fourier transform spectroscopy is now a widely-used tool for probing structure, dynamics and fluctuations that are obscured in linear optical spectroscopies, enabling new insights into the structure and function of complex molecules, aggregates and materials[1]. The approach borrows from early multidimensional Fourier transform nuclear magnetic resonance (NMR) spectroscopies[2]. The most commonly-used Fourier transform approaches use time-delayed pulse pair replicas created with interferometers. For multidimensional Fourier transform spectroscopic measurements at optical frequencies, this requires interferometric precision, with pathlength stabilities of $\sim\lambda/100$ to obtain reliable frequency axes and the separation of complex signal components with high signal-to-noise ratios[1]. To meet these challenges a variety of approaches have been employed[3], including passive[4-5] and active phase stabilization[6] in various geometries ranging from fully noncollinear[1,4-5] to partially[7-8] and fully collinear [9-11]. Other approaches have been developed to enable measurements in the rotating-frame that reduce the Nyquist frequency and sensitivity to phase instability. Here we demonstrate a phase-modulation-based approach to fluorescence-detected 2D Fourier transform spectroscopy using rapid-scanning and digital lock-in detection. The approach enables simultaneous detection of complex linear and multidimensional signals and can be readily integrated into a microscope for spatially-resolved measurements.

The phase-modulation approach was pioneered by Marcus and coworkers and used for wave packet interferometry [12] as well as fluorescence-detected [9-10] and photocurrent-detected [13] two-dimensional electronic spectroscopy (2DES). Their approach has been adopted by a number of other groups for fluorescence-detected 2DES [14-16] and other action-based 2D measurements [17-18].

The phase-modulation approach, which has also been termed dynamic phase-cycling [17], is closely related to phase-cycling which has been widely used in multidimensional spectroscopy [3, 7-8, 11, 19-22]. In phase-cycling, a pulse-shaper is typically used to create a pulse sequence with the desired inter-pulse delays and relative phases. Multiple measurements recorded with an appropriate set of delays and phases are then combined to isolate the signal of interest. In contrast, in the phase-modulation approach, an AOM imparts a distinct radio-frequency shift that modulates the pulse-to-pulse carrier-envelope phase shift. Linear and nonlinear spectroscopic signals of interest can then be isolated by detecting signals that are modulated at the appropriate frequency combination, either through lock-in detection [9, 12], super-heterodyne mixing [23] or direct digitization followed by signal selection via Fourier transform analysis [16]. Phase-modulation and phase-cycling provide an alternative to phase-matching approaches that are commonly used to enable "background free" detection of nonlinear signals. These approaches have the advantage that collinear pulses can be used, and since the coherent build-up of signal is not required, the sample size can be small compared to the excitation wavelength [19]. This makes these methods appealing for spatially-resolved measurements and studies of dilute samples.

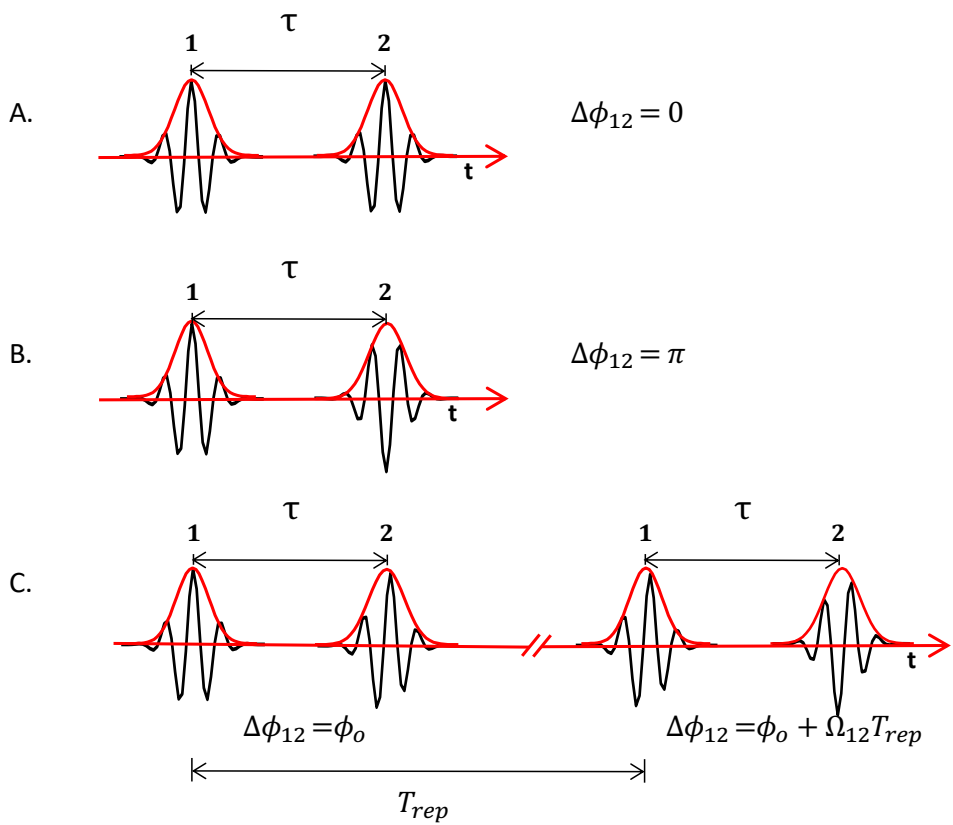

**Figure 1.** Pulse sequences used in Fourier transform spectroscopic measurements. A: Standard approach employing a pulse-pair generated by an interferometer, resulting in two time-delayed pulse replicas with identical carrier-envelope phase. B: Pulse pair generated by a pulse-shaper which can impart an arbitrary relative phase between the pulses (illustrated here with a π phase shift). C: Phase-modulated approach in which the pulse pair is generated by a Mach Zehnder interferometer with AOMs operating at distinct frequencies $\Omega_1$ and $\Omega_2$ in each arm, yielding a pulse pair with relative phase that is modulated at $\Omega_{12} = \Omega_1 - \Omega_2$. Thus the phase difference between the pulse pair changes dynamically for consecutive pulses in steps of $\Omega_{12}T_{rep}$ where $T_{rep}$ is the repetition rate of the laser.

The origins of phase-modulation-based Fourier transform spectroscopy can be traced back to the early work of Scherer et al. using phase-locked pulse pairs for fluorescence-detected wavepacket interferometry[24]. In their experiments a piezo-mounted mirror was used in combination with a phase-locked-loop to maintain a fixed relative phase between pulse pairs while varying the inter-pulse time delay. This step-scan approach effectively transforms the measurement into the rotating-frame, eliminating the high (optical) frequency modulation to isolate the rovibrational molecular dynamics in molecular iodine. In contrast to the piezo-driven mirror approach of requiring a fixed relative pulse-pair phase, the AOM-based method varies the relative pulse pair phase continuously at a well-defined frequency (see Figure 1), avoiding the need for phase-locking or stabilization and enabling rapid scanning. By operating in the rotating-frame, the AOM-based approach to multidimensional spectroscopy reduces the necessity for high phase-stability interferometers.

To date, most AOM-based approaches to multidimensional spectroscopy have used step-scanning and lock-in detection referenced to carefully constructed frequency references to extract the signals of interest [12,9-10,13,14-16,17-18]. Karki et al. demonstrated that an AOM-based phase-modulation combined with digital lock-in approach enables simultaneous detection of multiple multidimensional signals[16]. A continuous-scanning AOM-based approach was recently implemented by Autry et al. to simultaneously acquire multiple multidimensional spectra[23], employing a super-heterodyne mixing approach to isolate the signals of interest and a reference laser for interferometric tracking of the delays. Here we combine continuous-scanning with digital lock-in detection to acquire linear excitation, rephasing and nonrephasing fluorescence-detected 2D spectra simultaneously with interferometric tracking of the delays and accurate phasing in a single scan.

**Phase modulation and continuous scanning**

In the phase modulation approach to fluorescence-detected 2DES, the phases of the four collinear pulses are modulated by individual AOMs at four respective frequencies, $\Omega_1$, $\Omega_2$, $\Omega_3$, and $\Omega_4$. As a result, each pulse is tagged with a unique radio-frequency $\Omega_i$, which replaces the unique wavevectors, $\mathbf{k_i}$, of a standard fully noncollinear 2DES experiment[9-10]. In this case, the rephasing and non-rephasing signals arise from the four-wave mixing (FWM) populations which modulate at the linear combination of radio frequencies of the individual pulses. Particularly, the frequency combination $\Omega_R = -\Omega_1 + \Omega_2 + \Omega_3 - \Omega_4 = -\Omega_{12} +$ $\Omega_{34}$ and $\Omega_{NR} = \Omega_1 - \Omega_2 + \Omega_3 - \Omega_4 = \Omega_{12} + \Omega_{34}$ correspond to rephasing and non-rephasing signals, respectively, where $\Omega_{12} = \Omega_1 - \Omega_2$ and $\Omega_{34} = \Omega_3 - \Omega_4$.

In the implementation of Tekavec et al.[9] the rephasing and non-rephasing signals are detected in parallel via phase-sensitive lock-in detection with respect to appropriate reference signals to select the signals of interest while rejecting other oscillatory and stationary fluorescence signals. The reference signals for lock-in detection are constructed by recording spectrally-narrowed interference signals from MZ1 and MZ2, which modulate at $\Omega_{12}$ and $\Omega_{34}$ respectively. Through detection relative to the reference wavelength, the signal is physically undersampled, making the measurement insensitive to phase noise caused by mechanical delay fluctuations in the interferometer arms, thus avoiding the need for active-phase stabilization.

In contrast to the approach of Tekavec et al.[9], rather than lock-in detection, we take a digital signal acquisition approach in combination with rapid scanning. Upon Fourier transformation of the time domain fluorescence signal, the signals of interest can be distinguished by their distinct modulation frequencies imparted by the AOMs. In addition to digitizing the phase-modulated fluorescence signal, we simultaneously record several additional signals to enable interferometric determination of the time delays and accurate phasing of the data. In total, we simultaneously record four signals while continuous scanning of the $t_1$ and $t_3$ delays: the four-pulse fluorescence signal $S_{4PF}(t)$, the four-pulse reference signal $S_{4Pref}(t)$ and two mixer signals $M_{12}(t)$ and $M_{34}(t)$. The reference signal $S_{4Pref}(t)$, obtained after spectrally narrowing the four-pulse interference signal in a monochromator centered at frequency $\omega_{ref}$, is used for the interferometric determination of $t_1$ and $t_3$. The two mixer signals $M_{12}(t)$ and $M_{34}(t)$ measure $\Omega_{12}(t)$ and $\Omega_{34}(t)$ respectively and enable tracking of frequency drift in the AOMs. In addition to measuring the fluorescence-detected 2D spectrum, processing the linear components of $S_{4PF}(t)$ enables measurement of the linear fluorescence excitation spectrum, accurate determination of $t_1 = 0$ and $t_3 = 0$ and measurement of unknown spectral phase differences $\varphi_{12}(\omega)$ and $\varphi_{34}(\omega)$ which can arise due to small misalignment or pathlength differences through the AOMs in the two arms of MZ1 and MZ2 respectively. Details of the data processing for determination of the interferometric delays $t_1$ and $t_3$, as well as $t_1 = t_3 = 0$, $\varphi_{12}(\omega)$ and $\varphi_{34}(\omega)$ are given in Appendix A. These parameters are then used for the processing and accurate phasing of the fluorescence-detected 2DES signals.

**Experimental Methods:**

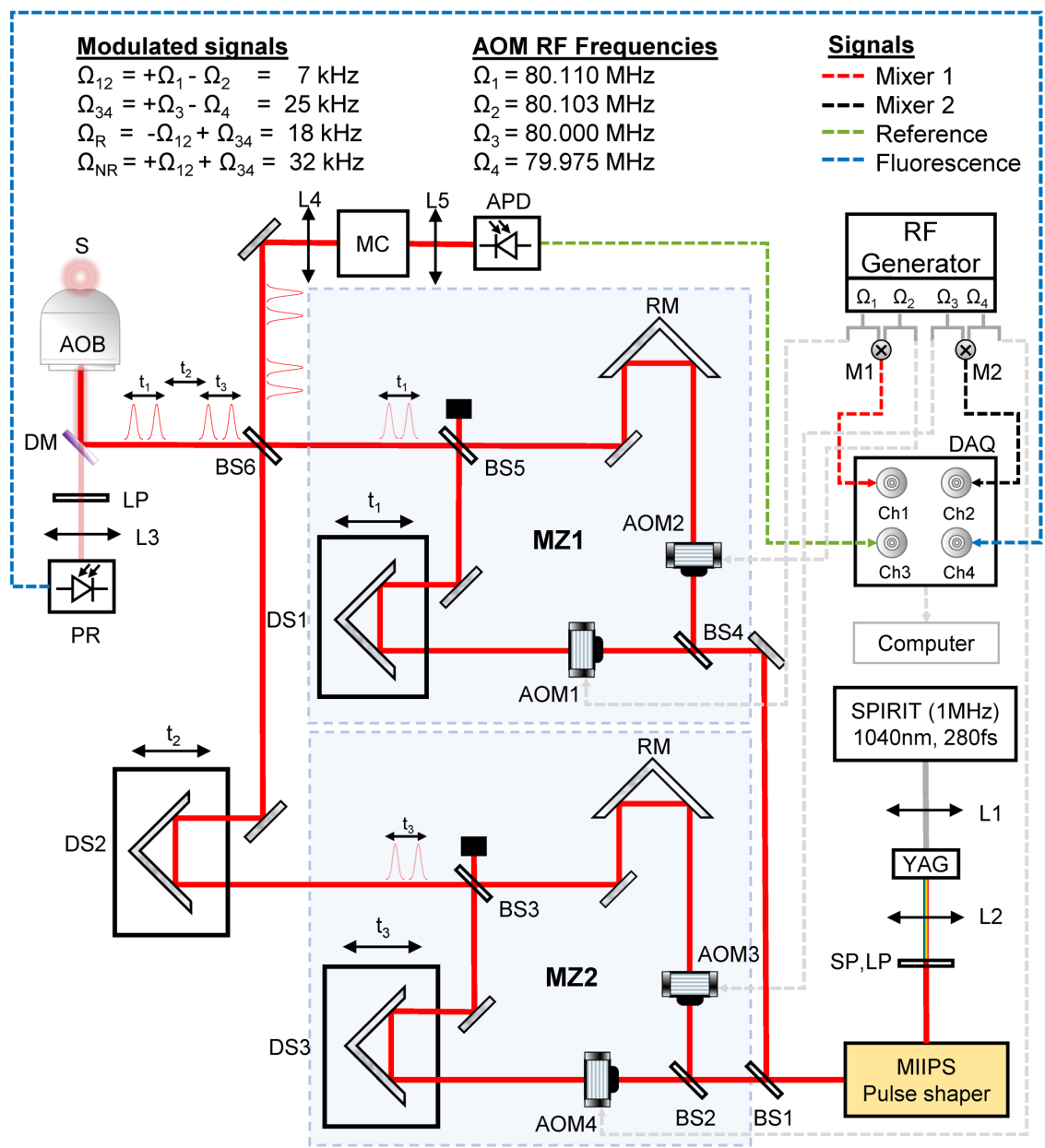

**Figure 2.** Experimental setup of the continuous-scanning phase-modulated fluorescence-detected 2D spectrometer (PM-F2DES). Beamsplitter (BS), Acousto-Optic Modulator (AOM), Mach-Zehnder interferometer (MZ), Radio frequency Mixer (M), Delay Stage (DS), Retroreflector Mirror (RM), Monochromator (MC), Avalanche Photodiode (APD), Variable Gain Photoreceiver (PR), Short-pass Optical Filter (SP), Long-pass Optical Filter (LP), Dichroic Mirror (DM), Sample (S), Air objective (AOB), Data acquisition card (DAQ).

The experimental setup, shown in Fig. 2, is based on the original design by Marcus and coworkers [12]. The pulse train from a 1 MHz Ytterbium amplified laser (Spectra Physics Spirit-HE) is focused in a 4mm thick Yttrium Aluminum Garnet crystal (YAG) using a 5cm focusing lens (L1), resulting in white-light generation (WLG). Re-collimation of the beam to a 3.7mm diameter ($1/e^2$) is obtained using a 5cm focal length lens (L2). A pair of short-pass (SP 950nm OD4, Edmund) and long pass optical filters (LP 650nm OD4, Edmund) are used in order to spectrally narrow the pulse and reject the residual fundamental light. To ensure that the pulses are Fourier-transform-limited at the sample position, the beam is routed to a SLM-based pulse shaper (MIIPS 640P, Bio-photonic Solutions) for dispersion pre-compensation and further bandwidth reduction. After the pulse shaper, a 50/50 beam splitter (BS1, Newport, 10B20BS.2) splits the pulse energy equally and directs it to two Mach-Zehnder interferometers (MZ1 & MZ2) in order to generate the excitation and detection pulse pairs.

At the entrance of each interferometer beam splitters (BS2 and BS4) divide the pulse energy equally along the MZ's fixed and moving arms. An acousto-optic modulator (AOM, Isomet, M1142-SF80L) inserted in the optical path of each arm, modulating at a particular radio frequency $\Omega_i$, imparts a time-varying phase shift to the consecutive pulses of the pulse train[12]. All four AOMs are driven by the same RF signal generator (Novatech, 409B) which uses a common internal clock for generating the modulation signals $\Omega_{1,2,3,4}$ = (80.110, 80.103, 80.000, 79.975) MHz. Although referenced to a common clock, small frequency drifts can result in phase errors observed when processing the data. Simultaneous recording of the waveforms used to drive the AOMs allows correction of such errors. To track the time-dependence of the difference frequencies Ω12 and Ω34, the difference frequency output of two frequency mixers (M1 & M2, Mini-Circuits, ZAD-1-1+) yield $M_{12}(t)$ and $M_{34}(t)$. Each pair of the modulated pulses is recombined on a beam splitter (BS3 & BS5) at the interferometer outputs, with the interpulse delay within each interferometer (t1 and t3), varied using a mechanical delay stage (DS, Newport, M-VP25XL). While the time delays t1 and t3 correspond to the coherence and detection times respectively, the waiting time t2 is controlled by an additional delay stage (DS2, Newport, ILS150). All three mechanical stages are controlled using a motion controller (Newport, XPS-Q4).

Finally, both pulse pairs are recombined to form a collinear four pulse train at BS6 with half of the energy used for interacting with the sample. The other half is used for detecting the spectrally-narrowed four-pulse reference signal $S_{4Pref}(t)$ for interferometric determination of the t1 and t3 delays. The spectral narrowing is achieved using a monochromator (MC, Dynacil, MC1-05G, 300μm slits resulting in 4.2nm FWHM bandwidth centered at 825nm) and the reference signal is detected via an avalanche photodiode (APD, Thorlabs, APD130A).

The four pulse sequence is directed to a stage-scanning inverted microscope (Olympus 1X51). A 875nm dichroic mirror (DM, Shemrock) reflects the incoming beam towards an air objective NA 0.6 (AOB, Olympus, LUCPlanFLN 40x) which focuses the beam in a flowcell of 200μm pathlength (Starna, 48-Q-0.2/UTWA-2). The emitted fluorescence is collected by the same objective and after being filtered by two long pass optical filters (AVR optics, TLP01-887 and Edmund optics, OD4-850 nm) is focused on a low-noise photoreceiver PR (FEMTO Messtechnik GmbH, OE-200-SI) using a 5cm focal length lens (L3).

For a fixed waiting time t2, the fluorescence-detected 2DES data is acquired by continuously scanning t3 back and forth at an average speed of 0.1 mm/s (and acceleration of 100 mm/s$^2$) between -80 fs and 80 fs, while slowly scanning t1 in a single direction from +80 fs and -80 fs at 0.0002 mm/s. All signals of interest ($S_{4PF}(t), S_{4Pref}(t), M_{12}(t)$ and $M_{34}(t)$) are digitized, recorded simultaneously and synchronized with our laser's repetition rate (1.0128334 MHz), using a 16-bit data acquisition card (DAQ, National Instruments, NI-USB 6366).

**Results**

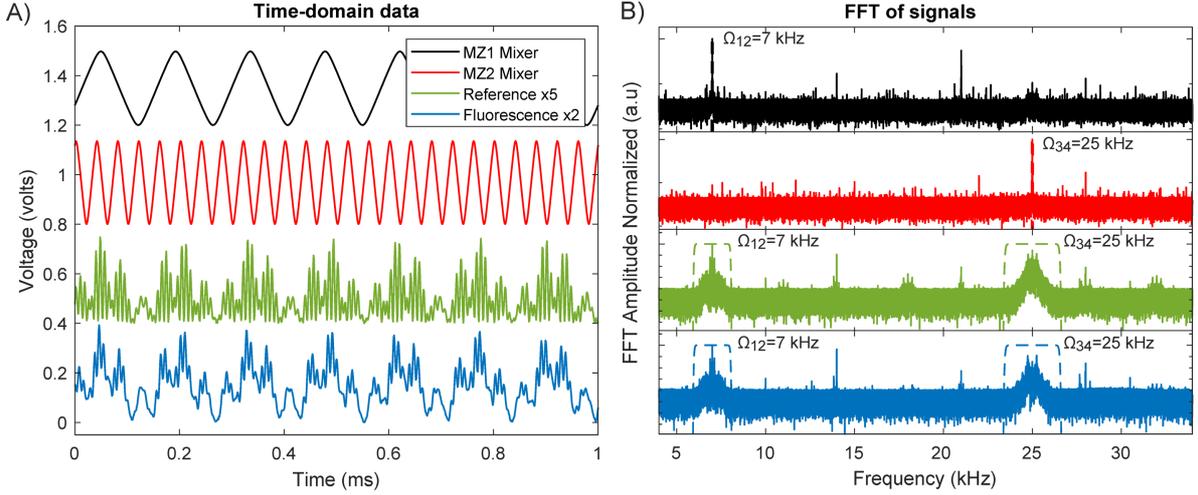

**Figure 3:** A: Time domain signals acquired for phase-modulated rapid-scanning fluorescence-detected 2DES. These signals are: mixer signals for MZ1 and MZ2, the four-pulse reference signal for interferometric tracking of the time delays and the four-pulse fluorescence signal. The signals are offset for clarity. (Note that a small subset of the full scan is shown). B) Fourier transform of the time domain signals shown in A, indicating the frequencies at which different signals appear. Also shown are the filter functions used to extract the quantities of interest.

Figure 3A shows the time domain signals of $M_{12}(t)$ and $M_{34}(t)$, $S_{4PF}(t)$ and $S_{4Pref}(t)$ that are simultaneously digitized, enabling measurement of the linear excitation spectrum, rephasing and non-rephasing fluorescence-detected 2DES signals. The Fourier transforms of these signals (taken from the full 2D dataset in which MZ1 and MZ2 are scanned as described above) are shown in Figure 3B. The Fourier transform of the mixer signals $M_{12}(t)$ and $M_{34}(t)$ display the expected peaks at the AOM difference frequencies $\Omega_{12}$ and $\Omega_{34}$ for MZ1 and MZ2 respectively. The Fourier transform of the fluorescence signal shows the strongest peaks at $\Omega_{12}$ and $\Omega_{34}$, arising from the linear excitation signal generated by the two-pulse sequence from MZ1 and MZ2 respectively. Similar peaks are also observed for the reference signal. We attribute the spectral broadening of these peaks to the fact that the time dependence of the time delay significantly differs from a linear variation, due to fluctuations of the path length in the interferometers. Due to the smaller amplitude of the nonlinear signals the Fourier transform of the fluorescence signal does not show clear peaks at $\Omega_R= -\Omega_{12}+\Omega_{34}$ and $\Omega_{NR}= \Omega_{12}+\Omega_{34}$ for the respective rephasing and non-rephasing signals prior to correction for these fluctuations. The retrieval of interferometric time delays, and the phasing of the fluorescence-detected 2D signals is achieved as detailed in Figure 4.

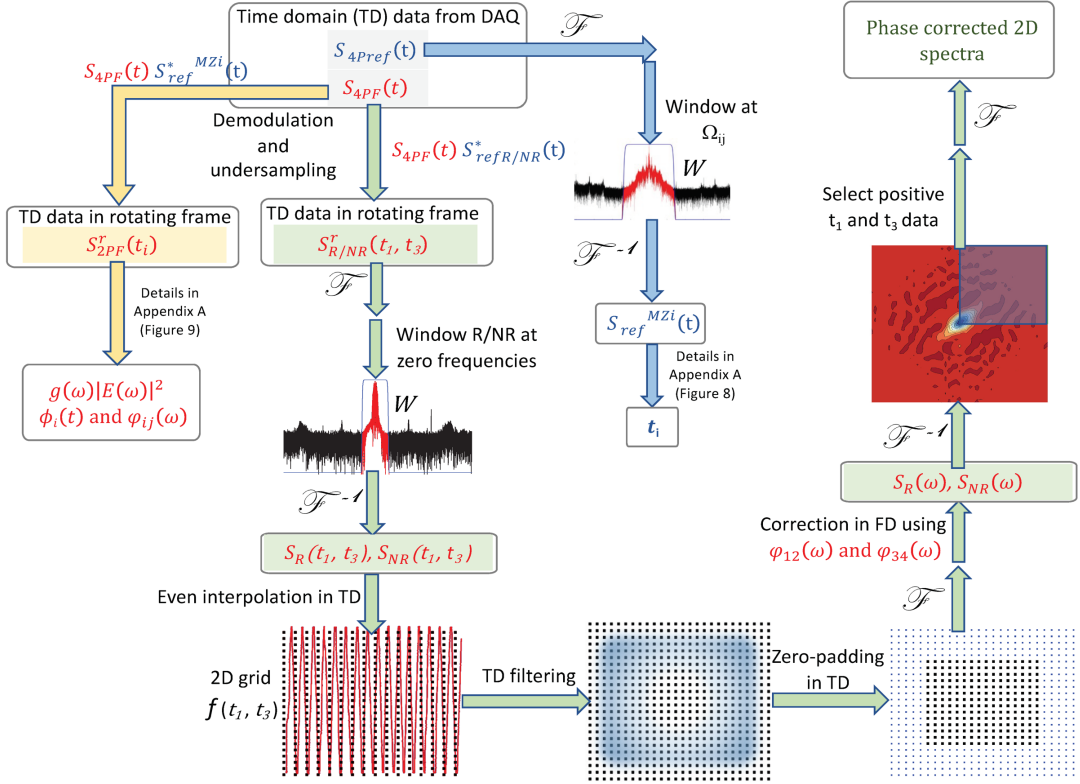

**Figure 4:** Flowchart depicting the processing of fluorescence data in order to obtain the phase-corrected fluorescence-detected 2D rephasing and nonrephasing spectra (green branch). The blue branch shows how the two MZ reference signals $S_{ref}^{MZ1}(t)$ and $S_{ref}^{MZ2}(t)$ and interferometric time delays $t_i$ can be extracted from the four-pulse reference signal $S_{4Pref}(t)$ (more details have been outlined in Appendix A, Figure 8). These two MZ references are used for the demodulation and undersampling of the fluorescence signal $S_{4PF}(t)$ to get $S_{2PF}^r(t_i)$, as shown in the yellow branch. Note that the two MZ reference signals are also used in appropriate combinations to construct the rephasing ($S_{refR}(t) = S^*{}_{ref}^{MZ1}(t)\, S_{ref}^{MZ2}(t)$) and nonrephasing reference signals ($S_{refNR}(t) = S_{ref}^{MZ1}(t)\, S_{ref}^{MZ2}(t)$). The reference signals $S_{refR/NR}(t)$ are then used to perform the demodulation of $S_{4PF}(t)$ to obtain $S_{R/NR}^r(t_1, t_3)$, shown in the green branch. Correction of the fluorescence signals for spectral phase imbalances in the MZs are made in the frequency domain using $\varphi_{12}(\omega)$ and $\varphi_{34}(\omega)$.

Finally, the inverse Fourier transform allows for selection of the correct time ordered signals, from which the phased fluorescence-detected signals are obtained upon Fourier transform.

*Determination of interferometric time delays*

To determine the interferometric time delays $t_1$ and $t_3$, the two-pulse reference signals $S_{ref}^{MZ1}(t)$, $S_{ref}^{MZ2}(t)$ are extracted via filtering of the four-pulse reference signal $S_{4Pref}(t)$ in the frequency-domain at $\Omega_{12}$ and $\Omega_{34}$ for MZ1 and MZ2 respectively. Figure 5A shows the Fourier transform of the $S_{4Pref}(t)$ signal, along with the filters of width 2 kHz and 3 kHz used for extracting $S_{ref}^{MZ1}(t)$ and $S_{ref}^{MZ2}(t)$ respectively. The extracted reference signals, along with the mixer signals $M_{12}(t)$ and $M_{34}(t)$ then enable interferometric determination of $t_1$ and $t_3$. Upon filtering, inverse Fourier transformation enables determination of the time domain signal phase. Demodulation of the $S_{ref}^{MZ1}(t)$ and $S_{ref}^{MZ2}(t)$ reference signals from the respective MZ1 and MZ2 mixer phase yields the interferometrically-determined $t_1$ and $t_3$ delays as shown in Figure 5B, where knowledge of $\omega_{ref}$ has been used. Further details of this procedure are given in Appendix A.

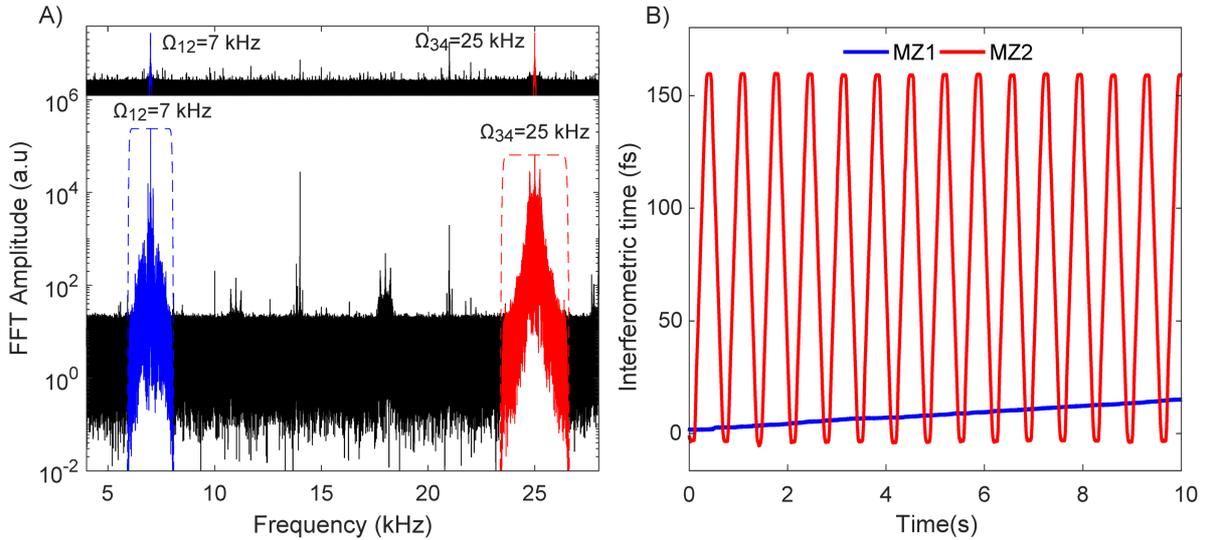

**Figure 5:** Interferometric determination of the time delays. A: The Fourier transform of the mixer signals (top) and four-pulse reference signal (bottom). Also shown are the filter functions used to extract the two-pulse reference signals $S_{ref}^{MZ1}(t)$, $S_{ref}^{MZ2}(t)$ for each MZ. B: Interferometrically-determined $t_1$ and $t_3$ delays, illustrating the unidirectional scanning of $t_1$

(MZ1) and the bidirectional scanning of t₃ (MZ2) (note that only a subset of the full scan is shown here).

*Determination of zero delays and spectral phase correction*

Upon Fourier transform of the four-pulse fluorescence signal $S_{4PF}(t)$ and filtering about the central frequencies $\Omega_{12}$ and $\Omega_{34}$ the two-pulse linear fluorescence signals $S_{2PF}^{MZ1}(t)$, $S_{2PF}^{MZ2}(t)$ can be isolated to enable measurement of the spectral phase difference $\varphi_{12}(\omega)$ and $\varphi_{34}(\omega)$ between the MZ pulse pairs, and precise determination of $t_1 = 0$ and $t_3=0$ as detailed in Appendix A. In Figure 6B (top panel) we show the linear fluorescence signal for MZ2, using the interferometrically-determined time axes, after demodulation with respect to the appropriate reference signal. Emphasis should be given to the high reproducibility of the data as 360 continuous repeated scans of MZ2 are overlayed in this figure. To obtain the zero delays and spectral phase differences for each interferometer, the data are first interpolated to evenly-spaced time points and the repeated MZ2 scans are averaged to optimize the signal-to-noise ratio. The Fourier transform then yields the measured $\varphi_{12}(\omega)$ and $\varphi_{34}(\omega)$, which arise from slight imbalances in the dispersion of MZ1 and MZ2 imparted by the AOMs. As detailed in Appendix A, the product of the fluorescence excitation spectrum $g(\omega)$ and the pulse power spectrum $|E(\omega)|^2$ can also be obtained from this measurement. These quantities are shown for both MZs in Figure 6. An independent measurement of the pulse power spectrum by interferometric autocorrelation could be used to determine $g(\omega)$ should that quantity be of interest.

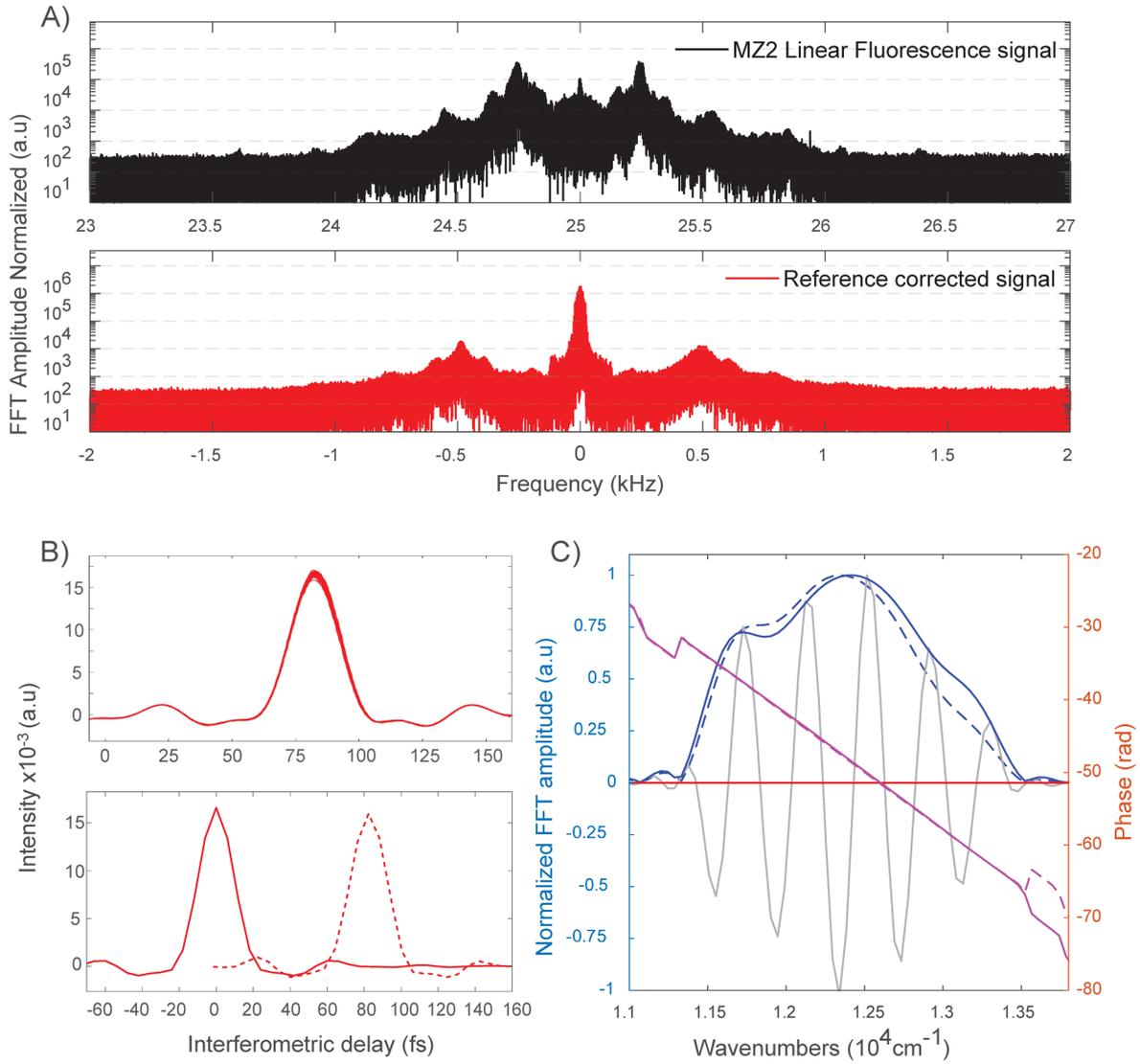

**Figure 6:** A: Top: Fourier transform of the linear fluorescence signal for MZ2 (black) showing distinct peaks due to the Doppler shift caused by the bidirectional scanning. Bottom: Fourier transform of the linear fluorescence signal for MZ2 after demodulation with respect to the reference signal (red). B: Top: Linear fluorescence signal for MZ2, shows the reproducibility of the multiple MZ2 scans. Bottom: Linear fluorescence signal for MZ2 before (dashed) and after (solid) correction of the spectral phase, yielding a symmetric scan about $t_3 = 0$. C: Real part (blue) and imaginary part (red) of $g(\omega)|E(\omega)|^2$ of IR140 measured by MZ1(dashed) and MZ2 (solid). Also shown is the Real part of $g(\omega)|E(\omega)|^2$ prior to correction (grey) of the spectral phase (magenta).

*Processing and phasing of fluorescence-detected 2D spectra*

As depicted in Figure 4, to obtain absorptive fluorescence-detected 2D spectra, the complex rephasing and non-rephasing signals are processed separately prior to addition. The rephasing and nonrephasing signals present in the four-pulse fluorescence signal $S_{4PF}(t)$ can be obtained after using the appropriate combinations of the two-pulse reference signals denoted $S^*_{refR/NR}(t)$. The isolated rephasing and nonrephasing signals are obtained upon filtering the already demodulated and stage corrected data in the frequency domain using filters of width 60 Hz centered at zero frequency. Using the interferometrically-determined $(t_1, t_3)$ values, the rephasing signal $S_R(t_1, t_3)$ and nonrephasing signal $S_{NR}(t_1, t_3)$ are interpolated to an evenly spaced grid of time points prior to Fourier transformation with respect to $t_1$ and $t_3$.

After flipping the rephasing spectrum along the $\omega_3$ dimension, both rephasing and non-rephasing spectra are phase corrected in the frequency domain upon subtracting $\varphi_{12}(\omega) + \varphi_{34}(\omega)$ as determined from the linear fluorescence signals (see Appendix A). Inverse Fourier transformation then allows selection of the time-domain quadrant corresponding to the correct scanning of the time delays $(t_1 > 0, t_3 > 0)$ for determination of the phased 2D signals via Fourier transform. Figure 7 shows the resulting fluorescence-detected nonrephasing, rephasing and absorptive 2D spectra at $t_2 = 0$.

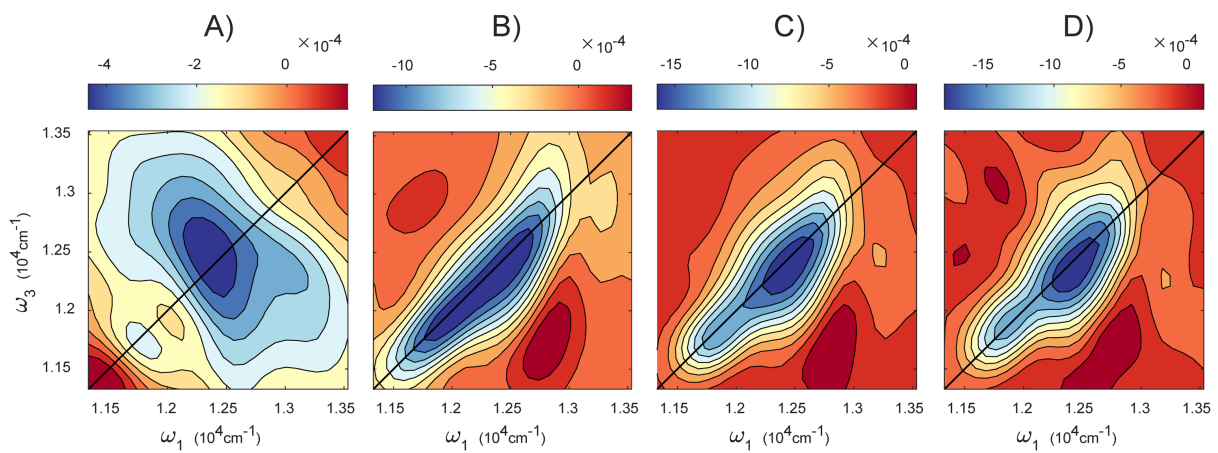

**Figure 7:** Fluorescence-detected 2D spectra for IR 140 in DMSO at $t_2$ = 0. A: Real nonrephasing spectrum, B: Real rephasing spectrum, C: Real absorptive spectrum (120 s acquisition time), D: Real absorptive spectrum (6 s acquisition time).

**Discussion**

Continuous-scanning offers the advantage of speed in comparison to step-scanning methods where acquisition times are increased due to settling times of the delay stages. Where used, active phase-locking can further increase acquisition times. Decreasing acquisition time is particularly important for fluorescence-detected 2DES due to photobleaching. This will be particularly important for measurements on small ensembles approaching the single molecule level.

The digital lock-in approach has the advantage of enabling the simultaneous acquisition of multiple multidimensional spectroscopic measurements. Here we have demonstrated the acquisition of linear fluorescence excitation spectra, as well as fluorescence-detected 2D rephasing and non-rephasing signals. We note that in typical implementations of 2DES, rephasing and nonrephasing spectra are recorded in separate scans, which can lead to difficulties in "phasing" the data[3]. In phase-modulated 2D measurements based on conventional lock-in detection, the phase is typically adjusted such that the real parts of the signal are maximized during the overlap of the pulses ($t_1 = t_3 = 0$).[9, 15] As discussed by Karki and coworkers, absolute signal phase determination may enable the separation of true nonlinear signals from those produced by incoherent mixing of linear signals that can occur in action-based 2D spectroscopy[25]. The simultaneous acquisition of two-quantum 2D spectra could be added to our approach, but would require digitization at considerably higher sampling rates to accommodate the higher frequency modulation of the two-quantum signals. We note that our approach is somewhat analogous to the recent demonstration of Mueller et al. in which they used a 125-fold phase-cycling scheme for the collection of multiple multidimensional signals within 8 minutes[26]. Their method employs step-scanning, and the pulse-shaper used to implement the experiment has different bandwidth and repetition rate considerations than our approach. The AOMs impart spatial chirp that may be important to compensate, depending on the application. Compensation could be achieved via prisms, the use of spatial filtering, or double-passing of the AOMs[27-28]. The phase-modulation approach has been used with laser sources varying in repetition rate from ~80MHz to 250 kHz. In order to effectively suppress 1/f noise, modulation rates should be selected to move the detection window to higher frequencies, within the digital sampling capabilities of the setup. We note that in contrast to the fully digital lock-in approach we adopt here, continuous-scanning with interferometric determination of the time delays could be achieved via conventional lock-in detection. To enable interferometric tracking of the delays, appropriate combinations of the MZ1 and MZ2 mixer signals would need to be constructed for use as references to isolate the signals of interest. In such an implementation, the number of signals that could be simultaneously detected would likely be limited by the number of channels on the lock-in amplifier.

While here we have collected fluorescence-detected linear and 2D spectra, an analogous digital lock-in phase-modulated approach could be readily developed for coherent 2D measurements, in which a fourth pulse would bypass the sample and be used for time domain heterodyne detection of the coherent 2D signal in a similar manner to the recent demonstration by Autry et al[23].

**Conclusion and perspectives**

In conclusion, we have presented continuous-scanning approach to fluorescence-detected two-dimensional electronic spectroscopy that combines acousto-optic phase-modulation with digital lock-in detection. We have demonstrated the method by measuring the linear excitation spectrum and absorptive fluorescence-detected two-dimensional electronic spectra of a laser dye. Advantages of phase-modulation include the suppression of 1/f laser noise due to the shifting of the signal detection window to frequencies above ~1 kHz where laser noise dominates. Digital lock-in detection offers a straightforward path to the simultaneous measurement of linear and nonlinear spectroscopic signals without the need for a multiple-channel lock-in amplifier. Our experimental setup enables straightforward interferometric tracking of the time delay to allow robust real-time determination of the absolute signal phase. We expect our approach to be particularly useful for reducing the acquisition time of spatially-resolved fluorescence-detected multidimensional spectroscopy measurements of samples that exhibit photobleaching [15].

**Appendix A: Linear Fourier-transform spectroscopy with phase-modulation and continuous-scanning**

We consider in this appendix a Fourier-transform spectroscopy measurement of a signal that is linear in the pulse intensity (two field interactions), such as a first-order autocorrelation, or the two-pulse fluorescence signal in the current experiment. For generality, and to allow for the case of a dispersion-unbalanced Mach-Zehnder (MZ) interferometer, we assume that the total electric field delivered by the MZ interferometer reads $E_1(t) + E_2(t - \tau)$, where $E_2(t)$ is a possibly-dispersed replica of $E_1(t)$. The quantity $t$ is the time, here at the femtosecond timescale, and $\tau$ is the time delay between pulses in the two distinct arms of the MZ. For a single laser shot, let us first consider the energy at the output of the MZ that would be measured using a perfect integrating photodetector. This quantity is given by

$$I_{MZ}(\tau) \propto \int_{-\infty}^{\infty} |E_1(t) + E_2(t - \tau)|^2 dt = const + S(\tau) + S(\tau)^* \qquad (1)$$

where

$$S(\tau) = \int_{-\infty}^{\infty} E_2^*(t - \tau) E_1(t) dt = \frac{1}{2\pi} \int_0^{\infty} E_2^*(\omega) E_1(\omega) e^{-i\omega\tau} d\omega \qquad (2)$$

is the correlation function, equal to the Fourier transform of $E_2^*(\omega) E_1(\omega)$ according to the convolution theorem. This expression can be generalized to any time-invariant signal linear in the pulse energy, i.e. bilinear in the electric field, provided that the non-instantaneity of the response is appropriately taken into account. The previous expression must then be replaced with

$$S(\tau) = \frac{1}{2\pi} \int_0^{\infty} g(\omega) E_2^*(\omega) E_1(\omega) e^{-i\omega\tau} d\omega \qquad (3)$$

where $g(\omega)$ is the frequency-domain function that describes the response of the system to the two electric-field interactions, one from each arm of the MZ. This function is known to be a real quantity, as a consequence of $S(-\tau) = S(\tau)$ in case of a balanced interferometer (i.e. when $E_1 = E_2$). We retrieve the case discussed above of a first order autocorrelation pulse measurement with an ideal broadband detector by assuming $g(\omega) = $ constant, whereas $g(\omega)$ should be replaced with the actual detector spectral response function for a non-ideal photodetector. In the case of the linear fluorescence signal $S_{2PF}(\tau)$, the response function $g(\omega)$ represents the fluorescence excitation spectrum.

Let us now consider a periodic train of laser shots centered at times $t_n$. Considering the fact that AOMs in arms 1 and 2 of the MZ modulate the pulse phases at frequencies $\Omega_1$ and $\Omega_2$ respectively, and that the diffracted ultrashort pulse samples the AOM phase at time $t_n$, we obtain :

$$S_{2PF}(\tau, t_n) \propto \frac{1}{2\pi} \int_{-\infty}^{\infty} g(\omega) E_2^*(\omega) e^{i\Omega_2 t_n} E_1(\omega) e^{-i\Omega_1 t_n} e^{-i\omega\tau} d\omega$$

$$= \frac{1}{2\pi} \int_{-\infty}^{\infty} g(\omega) E_2^*(\omega) E_1(\omega) e^{-i\Omega_{12} t_n} e^{-i\omega\tau} d\omega \qquad (4)$$

where $\Omega_{12} = \Omega_1 - \Omega_2$. Assuming a constant stage scanning speed of $v$, the time delay is sampled at the laser shot time according to the relation $\tau(t_n) = 2vt_n/c$. Finally, the measured signal reads

$$S_{2PF}(t) \propto \frac{1}{2\pi} \int_{-\infty}^{\infty} g(\omega) E_2^*(\omega) E_1(\omega) e^{-i\left(\Omega_{12} + \frac{2v\omega}{c}\right)t} d\omega \qquad (5)$$

where $t$ now represents the measurement time, here at the sub-microsecond time scale. It is important to stress that the above expression is valid only when the scanning speed is small enough so that the frequency extent associated with the modulated signal is much smaller than the electronic detection bandwidth. Furthermore, the above expression is an idealization assuming that the AOM frequencies and stage speed are perfectly constant. In practice, we must take into account a small drift $\delta\Omega_{12}(t)$ of the AOM frequency difference. Even more important, due to mechanical imperfection, the stage speed is not exactly constant. The actual signal thus reads

$$S_{2PF}(t) \propto \frac{1}{2\pi} \int_{-\infty}^{\infty} g(\omega) E_2^*(\omega) E_1(\omega) e^{-i(\Omega_{12} t + \phi(t) + \omega\tau(t))} d\omega \qquad (6)$$

where

$$\phi(t) = \int_0^t \delta\Omega_{12}(t')dt' \tag{7}$$

is the accumulated AOM phase drift and

$$\tau(t) = \int_0^t \frac{2v(t')}{c} dt' \tag{8}$$

is the actual time delay, which can also include any contribution from vibrations in the interferometer. In order to measure these quantities with the required accuracy, we simultaneously record the mixer signal

$$M_{12}(t) \propto e^{-i(\Omega_{12}t+\phi(t))} \tag{9}$$

and a reference signal monitoring the interference of pulses 1 and 2 in a monochromator centered at frequency $\omega_{ref}$. The measured signal, again linear in the pulse energy, is consequently given by

$$S_{ref}(t) \propto \frac{1}{2\pi}\int_{-\infty}^{\infty} f(\omega)\, E_2^*(\omega)E_1(\omega)e^{-i(\Omega_{12}t+\phi(t)+\omega\tau(t))}d\omega \propto e^{-i(\Omega_{12}t+\phi(t)+\omega_{ref}\tau(t))} \tag{10}$$

where $f(\omega)$ is the monochromator spectral response, assumed to be a narrow peak centered on frequency $\omega_{ref}$. Selecting $\Omega_{12}$ via windowing in the frequency domain, we take into account any phase drift by directly subtracting point-by-point in the time domain:

$$S_{ref}(t)M_{12}^*(t) \propto e^{-i\omega_{ref}\tau(t)} \tag{11}$$

which yields the actual time delay $\tau(t)$ after phase unwrapping, allowing to average each measured data point in the appropriate time bin and thus record $S_{2PF}(\tau)$. The procedure for determining time delay $\tau(t)$ is depicted in Figure 8.

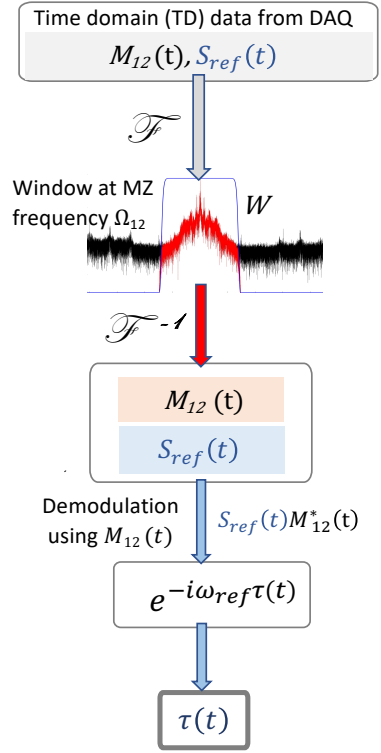

**Figure 8:** Flowchart depicting the processing of the time-domain data to obtain the interferometric time delay $\tau(t)$.

Furthermore, the signal of interest can be corrected with respect to AOM frequency drift and shifted to the rotating frame:

$$S_{2PF}^r(\tau) = S_{2PF}(t)S_{ref}^*(t) \propto \frac{1}{2\pi}\int_{-\infty}^{\infty} g(\omega)E_2^*(\omega)E_1(\omega)e^{-i(\omega-\omega_{ref})\tau}d\omega \quad (12)$$

We further assume identical spectra for the two pulses ($|E_1(\omega)| = |E_2(\omega)| = |E(\omega)|$) but allow for a possible spectral phase difference $\varphi_{12}(\omega) = \varphi_1(\omega) - \varphi_2(\omega)$ induced by different dispersion between arms 1 and 2 due to AOM differences or slight misalignment.

$$S_{2PF}^r(\tau) \propto \frac{1}{2\pi}\int_{-\infty}^{\infty} g(\omega)|E(\omega)|^2 e^{i(\varphi_{12}(\omega)-(\omega-\omega_{ref})\tau)}d\omega \quad (13)$$

Since $g(\omega)$ is purely real, the FT of $S_{2PF}^r(\tau)$ yields $g(\omega)|E(\omega)|^2$, the unknown spectral phase $\varphi_{12}(\omega)$ and the location of $\tau = 0$. As a time shift is associated with a linear slope, the spectral phase is defined such that its first-order derivative at center frequency is zero, so that the zero time delay can be defined unambiguously. The data processing steps are depicted in Figure 9.

Furthermore, an independent measurement of the power spectrum $|E(\omega)|^2$ by interferometric autocorrelation or other means enables determination of $g(\omega)$, the fluorescence excitation spectrum. In summary, the simultaneous measurements of $S_{2PF}(t)$, $S_{ref}(t)$ and $M_{12}(t)$ yield the interferometric time axis $\tau$ including $\tau = 0$, $\phi(t)$ and $\varphi_{12}(\omega)$.

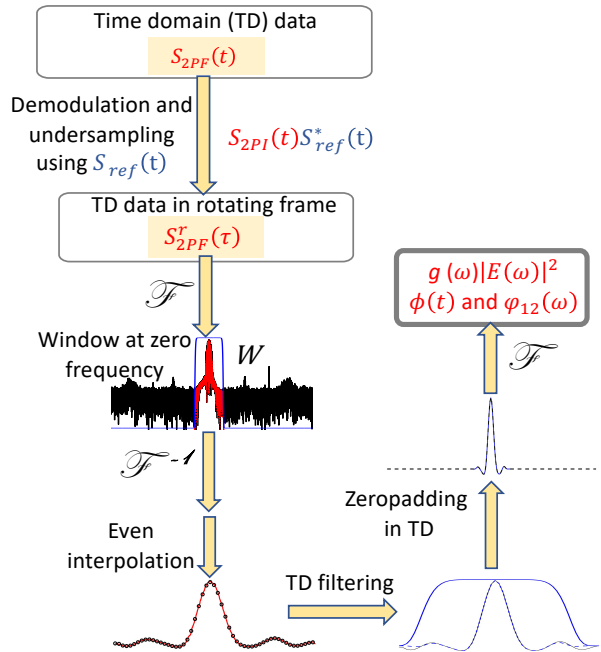

Figure 9: Flowchart depicting the processing of the linear two-pulse interference time-domain data to obtain $g(\omega)|E(\omega)|^2$, the unknown spectral phase $\varphi_{12}(\omega)$, $\phi(t)$ and the location of $\tau = 0$.

ACKNOWLEDGMENT

J.P.O., D. A. and A. J. acknowledge support from the AFOSR Biophysics program for support of this research under grant FA9550-15-1-0210 and FA9550-18-1-0124. J.P.O also acknowledges a visiting scientist grant from the Ecole Polytechnique.

DATA AVAILABILITY

The data that support the findings of this study are available from the corresponding author upon reasonable request.

**The following article has been submitted to/accepted by The Journal of Chemical Physics.**

**After it is published, it will be found at [Link](Link).**